# Spatially Inhomogeneous Bimodal Inherent Structure in Simulated Liquid Water

K. T. Wikfeldt[1], A. Nilsson[1,2] and L. G. M. Pettersson[1]

[1]Department of Physics, AlbaNova, Stockholm University, S-106 91 Stockholm, Sweden

[2]Stanford Synchrotron Radiation Lightsource, P.O.B 20450, Stanford, CA 94309, USA

ABSTRACT

In the supercooled regime at elevated pressure two forms of liquid water, high-density (HDL) and low-density (LDL), have been proposed to be separated by a coexistence line ending at a critical point, but a connection to ambient conditions has been lacking. Here we perform large-scale molecular dynamics simulations and demonstrate that the underlying potential energy surface gives a strictly bimodal characterization of the molecules at all temperatures as spatially inhomogeneous either LDL- or HDL-like with a 3:1 predominance for HDL at ambient conditions. The Widom line, indicating maximum fluctuations, coincides with a 1:1 distribution. Our results indicate a unified description of liquid water covering supercooled to ambient conditions in agreement with recent x-ray spectroscopy and scattering data.

**Introduction**

Water is the most common molecular substance and also the most unusual due to the number of anomalies in its thermodynamic properties. Water exhibits, *e.g.*, increased density upon melting and density maximum at 4°C both of which are related to the thermal expansion coefficient $\alpha_P$ which becomes negative for $H_2O$ water below 4°C. The isothermal compressibility $\kappa_T$ increases anomalously upon cooling below 46°C and so does the heat capacity at constant pressure $C_P$ below 35°C. All three thermodynamic response functions are related to fluctuations in the liquid ($\kappa_T$ to density fluctuations, $C_P$ to fluctuations in the entropy and $\alpha_P$ to their crosscorrelation) which *increase* upon cooling below a certain temperature. In the supercooled region below the freezing point these anomalous properties of water show indications of a power law divergence [1-3], but they clearly affect also ambient water properties where most processes of chemical and biological relevance occur. Many scenarios have been suggested to explain this anomalous behavior, *e.g.*, the critical-point free [4], singularity-free [5], stability-limit [6] and glass-transition [7] scenarios. At present, however, the liquid-liquid critical point (LLCP) scenario[8] appears to have the largest theoretical and experimental support[2,3,8-14] although no conclusive evidence has so far been given.

In the LLCP scenario two liquid phases, high density (HDL) and low density (LDL) liquid, are assumed separated at elevated pressure and supercooled temperatures by a coexistence line terminating at a critical point where thermodynamic response functions diverge. In the one-phase region, thermodynamic response functions show maxima in the *P-T* plane defining the so called Widom line as an extension of the coexistence line beyond the critical point. The Widom line divides the one-phase region into regions dominated by properties characteristic of either phase, but the distinction, and magnitude of the maxima, diminishes with increasing distance to the critical point[15].

Characterizing a phase transition requires an order parameter that in a relevant way describes the change in the system. HDL is characterized by the presence of interstitial molecules between the first and second hydration shells, leading to perturbed hydrogen-bonding and higher density; this is seen as smearing out of the 2nd shell at 4.5 Å. LDL exhibits more ice-like local order with well-separated first and second shells [16]. This distinction was recently employed by considering the distribution, $g_5(r)$, of the fifth neighbor finding a distinct bimodality and mixture-like behavior in the vicinity of the liquid-liquid transition in the model[17]. A more elaborate structural parameter is given by the local structure index, LSI,

[18,19] which reflects the order in both the first and second coordination shells. In simulations of water LSI gives a unimodal distribution peaking at low values (HDL-like) and extending to high values (LDL-like). In recent studies the LSI has instead been applied to the *inherent* structure in simulations of supercooled and ambient SPC/E water finding a bimodal distribution of low and high values in support of a two-state picture of water[20,21]; the inherent structure is obtained by quenching the instantaneous structure into the nearest local minimum on the underlying potential energy surface of the simulation, thus removing effects of thermal excitations[22,23].

Here we apply the LSI to the inherent structure in large-scale molecular dynamics simulations of TIP4P/2005[24] water and show that it gives a bimodal HDL and LDL distribution at all temperatures ranging from hot to supercooled with relative weight dependent on temperature and pressure. Specifically, at ambient conditions we find a 3:1 ratio with dominance of HDL consistent with recent x-ray spectroscopic results [25-28]. A striking result is that the reported Widom line (at subcritical pressures) and the liquid-liquid coexistence line (at high pressure) of the TIP4P/2005 model coincide closely with a 1:1 distribution between HDL- and LDL-like molecular species indicating maximal fluctuations and a facile way to find this important state point. We show that molecules classified as HDL-like and LDL-like based on their LSI values in the inherent structure are strongly spatially correlated and heterogeneously distributed also in the real, thermally excited structure. We also show that the distributions of local densities for the two species are displaced by about 0.05 g/cm$^3$, providing a natural interpretation of the isothermal compressibility anomaly and experimentally observed enhanced small-angle scattering at ambient and supercooled temperatures [28-32].

**Methods**

We perform classical MD simulations using the TIP4P/2005 force-field [24] with 45,000 molecules in the isothermal-isobaric (NPT) ensemble. Equilibration at supercooled temperatures is slow and the density and structural relaxation times increase quickly upon supercooling; the simulations reported here have thus been run for up to 45 ns at the lowest temperatures. Nosé-Hoover thermostats [33,34] and Parrinello-Rahman barostats [35] were used to control the temperature and pressure, respectively, and a time-step of 2 fs was applied (4 fs at the lowest temperatures). Long-range electrostatic interactions were treated with the particle-mesh Ewald method [36], long-range Lennard-Jones corrections are included and the TIP4P/2005 intramolecular geometry was constrained using the LINCS algorithm. Gromacs

4.0 [37] has been used on a parallel platform for all simulations. Further details of the simulations can be found in ref. [38] and its associated Supplementary Material.

The local-structure-index (LSI) [18,19] is defined for each molecule $i$ by ordering the nearest neighbors $j$ according to increasing distance to reference molecule $i$ as $r_1 < r_2 < r_3 < \cdots < r_{n(i)} < 3.7$ Å $< r_{n(i)+1}$ where $n(i)$ is the number of molecules that are within 3.7 Å from molecule $i$ (oxygen atom positions are used). The LSI distinguishes molecules with well separated first and second coordination shells from molecules with disordered environment, containing molecules in interstitial positions, through the parameter $I(i)$ defined by

$$I(i) = \frac{1}{n(i)} \sum_{j=1}^{n(i)} [\Delta(j;i) - \overline{\Delta}(i)]^2 \qquad (1)$$

Here $\Delta(j;i) = r_{j+1} - r_j$ and $\overline{\Delta}(i)$ is the average of $\Delta(j;i)$ over all neighbors $j$ of molecule $i$ within the cutoff. A low LSI corresponds to a disordered local environment (HDL) while a high LSI indicates a highly structured, tetrahedral coordination (LDL).

**Results and Discussion**

In figure 1 we show the evolution with temperature of the probability distribution of LSI values, $I(i)$, computed for inherent structures (IS) at different isobars. A clearly bimodal character is observed at all state-points. The stability of the minimum at 0.13-0.14 Å$^2$ at all temperatures and pressures is quite remarkable and a strong indication of a clear distinction between the two classes of local environments in the inherent structure. This implies that the 3N-dimensional potential energy surface (PES) on which the simulation evolves contains two qualitatively different types of local projections relating to *local* configurations of molecules in agreement with recent experimental x-ray absorption (XAS)[26,27,39] and emission (XES)[25,28] studies.

At 1, 1000 and 1500 bar there are sharp changes in the relative intensity of the high-LSI and low-LSI peaks at deeply supercooled temperatures, while at 2000 bar the temperature dependence is much weaker. It is also clear that at high temperatures the simulations sample very similar regions of the PES since only weak temperature dependence is seen. A strikingly clear connection to thermodynamic behavior is established in figure 2 where we show the populations in each class as function of temperature and pressure using the minimum around

0.13 Å$^2$ as classification. The populations at ambient conditions, ~25% high-LSI (LDL-like) and 75% low-LSI (HDL-like), coincide closely to experimental estimates from XAS[26,39] and XES[25,28], as well as from decomposition of the OH stretch region of the infrared spectrum analyzed in connection with the appearance of a fractional Stokes-Einstein relation in supercooled water[13].

Earlier studies[38,40,41] have located the LLCP and the Widom line of the TIP4P/2005 model where the LLCP was found close to P$_c$=1350 bar and T$_c$=193 K[40]. Figure 2 shows that both the Widom line below P$_c$ and the liquid-liquid phase transition (LLPT) above P$_c$ coincide closely with the temperatures where a 1:1 ratio between low-LSI and high-LSI populations is seen. The rate of change from low-LSI to high-LSI becomes greater with increasing pressure as the LLCP is approached, and we note in particular that the very steep slope at 1500 bar as T=190 K is approached is consistent with the possibility that a discontinuous jump would occur at the LLPT; here we have however been unable to directly observe the LLPT due to the insurmountably slow relaxation.

That the temperature dependence at 2000 bar becomes very weak fits well with the picture presented here since, with the LLCP of TIP4P/2005 water at 1350 bar as reported in ref.[40], the liquid has been pushed beyond the LLCP on the HDL side of the LLPT and should therefore display almost pure HDL behavior. The small but persistent population of high-LSI species at 2000 bar implies however that LDL-like fluctuations occur even at this pressure. Another simple, yet important, observation from figure 2 is the pressure dependence at isothermal conditions. Application of pressure shifts the balance from high- to low-LSI, consistent with the expected conversion of LDL- to HDL-like coordination with increasing pressure[16,42].

We now turn to an analysis of the low- and high-LSI species in the real, thermally excited structure (RS), focusing on ambient conditions, *i.e.* T=298 K and P=1 bar. Since the bimodal LSI distribution in the IS can be considered to reflect an underlying bimodality in the RS, which however is smeared out by thermal excitations in the simulation, we categorize molecules in the RS according to their LSI values in the IS. A convenient nomenclature for this scheme is to refer to *inherently* low- and high-LSI species, respectively, where we use as cut-off value the temperature and pressure invariant minimum at *I*(*i*)=0.13 Å$^2$.

Figure 3A shows oxygen-oxygen (O-O) pair-correlation functions (PCFs) separated into contributions from inherently low- and high-LSI species, respectively. Even though the assignment of subspecies is performed based on the IS and the effect of temperature masks the underlying bimodality in the RS, it is clearly seen that inherently low- and high-LSI species exhibit very different local environments in the simulation. Inherently high-LSI molecules have well defined coordination shells and a deep minimum between the first and second shells while inherently low-LSI molecules have somewhat weaker first-shell correlations and the higher shell structure is strongly diminished. Another way to decompose the radial correlations (figure 3B) is to calculate partial O-O PCFs for the two subspecies along with their cross correlation, where the former is the PCF between molecules of the same kind (*either* low- or high-LSI) and the latter is the PCF *between* the two classes. Here the differences are even more accentuated in that high-LSI species have significantly stronger first-shell correlations to other high-LSI species while the higher shell correlations between low-LSI species is further washed out. A shift to shorter distances of the second and third peaks is seen for the inherently low-LSI species both in figure 3A and 3B, an effect analogous to the effect of pressure on the O-O PCF [16,42]. Indeed, the partial O-O PCFs and structure factors, $S(k)$, of low- and high-LSI species in figures 3B and 3C show an overall qualitative similarity to the corresponding experimental counterparts for high- and low-density amorphous ice, respectively, obtained from x-ray and neutron diffraction [43,44].

Further information is contained in the PCF cross correlation between the two subspecies in figure 3B where the relatively lower amplitude of the cross correlation indicates that inherently low- and high-LSI species coordinate preferentially with species of the same kind, damping the cross correlation. Such a clustering effect is easily detected in the small-angle region of the partial structure factors, which we calculate as

$$S_{OO}^{\alpha\alpha}(k) = 1 + 4\pi\rho_\alpha \int r^2 \left(g_{OO}^{\alpha\alpha}(r) - 1\right) \frac{\sin(kr)}{kr} dr$$

and show in figure 3C. The index α refers to either inherently low- or high-LSI species and we use the respective partial number densities $\rho_\alpha$ of the two subspecies.

Pronounced small-angle scattering intensity is indeed seen - a clear signature of a spatially inhomogeneous distribution of inherently low- and high-LSI species; a random selection of molecules from the box gives no excess enhancement at low $k$. Note that the partial structure factor for the cross-correlations (not shown) becomes negative at low $k$, reflecting the spatial separation between the two species. As a result the three partial correlations, *i.e.* two intra-species and one inter-species contribution, add up to a total structure factor which displays a rather weak but measurable small-angle enhancement at ambient temperature[28,30,38].

The inhomogeneous distribution of the two species can be seen already by visually inspecting the simulation boxes. Figure 4 shows snapshots from the real structures at selected temperatures, where inherently low- and high-LSI species have been colored differently. At 230 K, close to the Widom line of the model, a very pronounced clustering effect is observed which was shown in ref. [38] to give a maximum in the small-angle scattering; the clustering effect can, however, be seen even at ambient (298 K) and warm (340 K) temperatures.

In figure 5 we show the density distribution in the simulated liquid subdivided into contributions from inherently low- and high-LSI species where the volume occupied by each molecule is obtained by constructing its Voronoi polyhedron in the RS. We observe an overall unimodal density distribution, as expected for a one-phase liquid, but it is decomposed by the LSI into two distributions differing in average density with inherently high-LSI species having lower local density. A natural interpretation thus emerges of the enhanced small angle scattering intensity at ambient temperature and below, observed experimentally [28-30,32] and in simulations [38], where the scattering contrast originates from an increasing density difference (0.047 g/cm$^3$ at 320 K, 0.056 g/cm$^3$ at 230 K) between the two species upon cooling to lower temperatures accompanied by an increasing population and spatial extent of the instantaneous low-density, LDL-like, high-LSI species. As seen in the inset of figure 4, when heating the liquid the average density of LDL-like species decreases somewhat less than for the HDL-like species. This is in accordance with the interpretation of recent x-ray spectroscopic and scattering data where the shift towards gasphase with increasing temperature of the features assigned to the disordered, HDL-like species (in XAS the pre- and main-edge[26,27,39,45], in XES the high emission energy peak[25,46]) has been interpreted as thermal excitation and expansion of this species while features assigned to tetrahedral, LDL-like species loose intensity but remain at fixed energy[25,28,39,45]. It is also consistent with the shift with temperature towards lower LSI values of the HDL-like component seen in fig. 1, indicating a more rapidly increasing disorder with increasing temperature for this component.

**Summary and Discussion**

We have shown that the local-structure index (LSI) applied to the inherent structure in simulations, here using the TIP4P/2005 force-field[24], reveals a structural bimodality in the underlying potential energy surface on which the thermal motion evolves. Analyzing separately the two distributions we find that high-LSI species exhibit a highly structured O-O pair-correlation function, as expected for LDL, while that of species with low LSI is similar to expectation for HDL with a partially collapsed second hydration shell. Counting the number of molecules in each distribution as function of temperature and pressure we find a strong dependence with HDL dominating 3:1 at ambient conditions while the state points with 1:1 population coincide with the Widom and coexistence lines in the model where indeed fluctuations are expected to be maximal.

Based on Voronoi tesselation in the real structure, but with assignment based on the inherent structure, the density distributions of high- and low-LSI species are found to be shifted relative each other. The difference in mean local density is found to depend on temperature decreasing from 0.056 g/cm$^3$ at 230 K to 0.047 g/cm$^3$ at 320 K, with the HDL-like component expanding above 253 K even though the density maximum around 277 K is accurately described by the simulation model[24]. This shows that with increasing temperature the higher-density, low-LSI component becomes more thermally excited, and expands relative the high-LSI component; this, together with the conversion of high-LSI species to low-LSI with increasing temperature indicates that high-LSI species are of lower enthalpy while low-LSI species contribute more to the entropy[28,47].

Recent x-ray absorption (XAS)[26,27,39] and emission (XES)[25,28,46] spectroscopy studies of ambient to hot water, as well as small-angle x-ray scattering (SAXS) studies of ambient to supercooled water[28,30], have been interpreted in terms of temperature-dependent fluctuations between strongly tetrahedral (LDL-like) and hydrogen-bond disordered (HDL-like) species, with the clearest indication of a structural bimodality from XES which shows two sharp and well separated peaks which interconvert but do not broaden with temperature. Although still debated[31,48-51] (see discussion in refs. [39,52,53]) this interpretation seems to find support in the present large-scale simulations, but only in the inherent structure, *i.e.* only when the effects of thermal motion and disorder are eliminated. How can this surprising observation be understood?

Returning to the initially discussed fluctuation-dependent thermodynamical properties ($\kappa_T$, $C_P$, $\alpha_P$) we note that commonly employed force-fields (SPC/E, TIP4P, TIP5P, TIP4P/2005) all give too high isothermal compressibility $\kappa_T$ compared to experiment at temperatures above 300 K while below 280 K $\kappa_T$ is instead underestimated [54]. Assuming that the experimental minimum in $\kappa_T$ at 46 °C indicates a switch between dominance of structural fluctuations below 46 ˚C and normal thermal motion above 46 ˚C, this would indicate that at low temperatures structural HDL to LDL fluctuations are underestimated, while on the other hand at high temperatures stochastic thermal motion is overestimated in the simulations. Both these effects would contribute to smearing out the underlying bimodality evident in the inherent structure and indicated by x-ray spectroscopy and scattering experiments. We speculate that a more accurate reproduction of fluctuation-dependent thermodynamical properties at all temperatures is required to retain aspects of the inherent bimodality also in the real simulated structure.

**Acknowledgements**

This work was supported by the Swedish Research Council (VR) and the National Science Foundation (US) CHE-080932. The MD simulations were performed on resources provided by the Swedish National Infrastructure for Computing (SNIC) at the NSC and HPC2N centers. Portions of this research were carried out at the Stanford Synchrotron Radiation Lightsource, a national user facility operated by Stanford University on behalf of the U.S. Department of Energy, Office of Basic Energy Sciences.

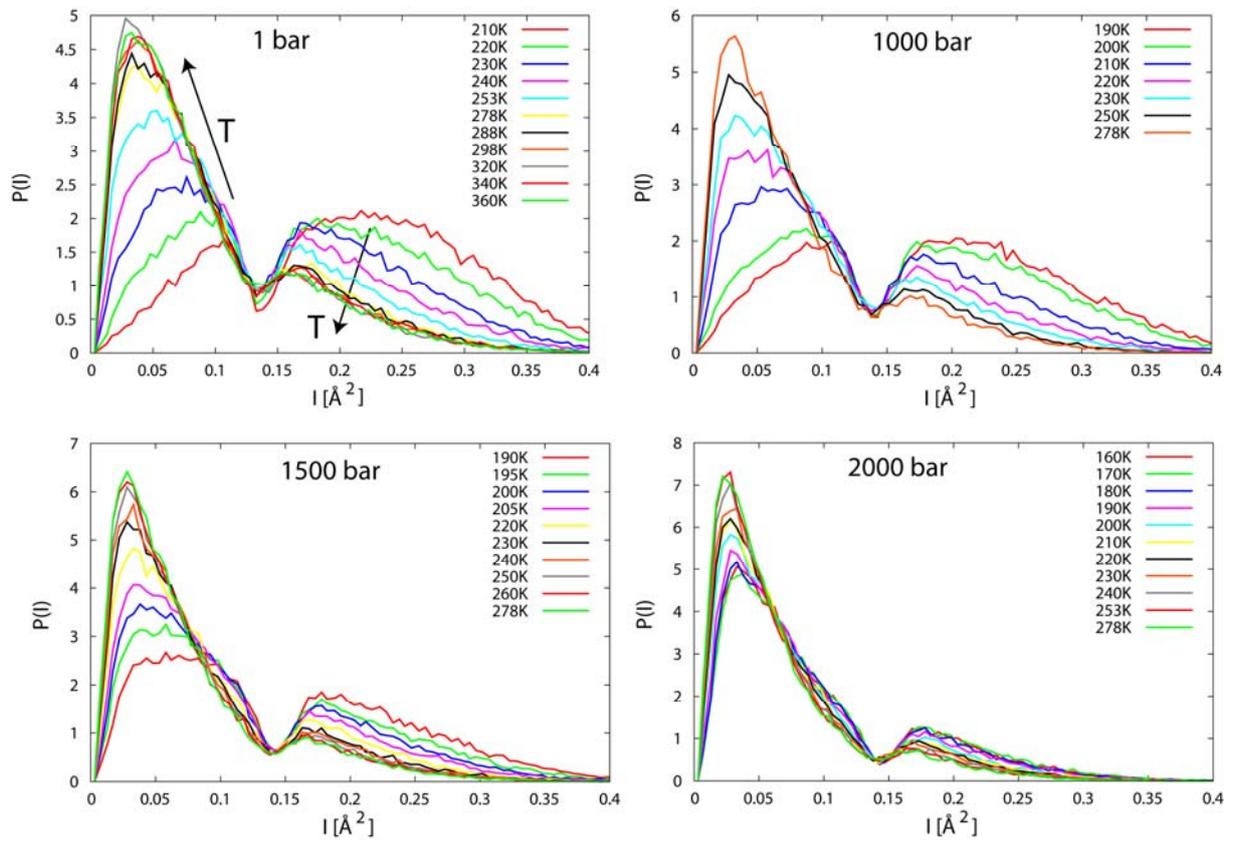

**Fig. 1.** Distributions of the local structure index parameter $I(i)$ obtained from energy-minimized "inherent structures" in simulations at ambient and elevated pressures, shown as function of temperature. Temperature and pressure invariant isosbestic points are seen around $I=0.13$-$0.14$ Å$^2$ at all pressures. Rapid changes are seen close to the Widom temperatures at 1 and 1000 bar and close to the liquid-liquid phase transition temperature at 1500. At 2000 bar the temperature dependence is strongly diminished.

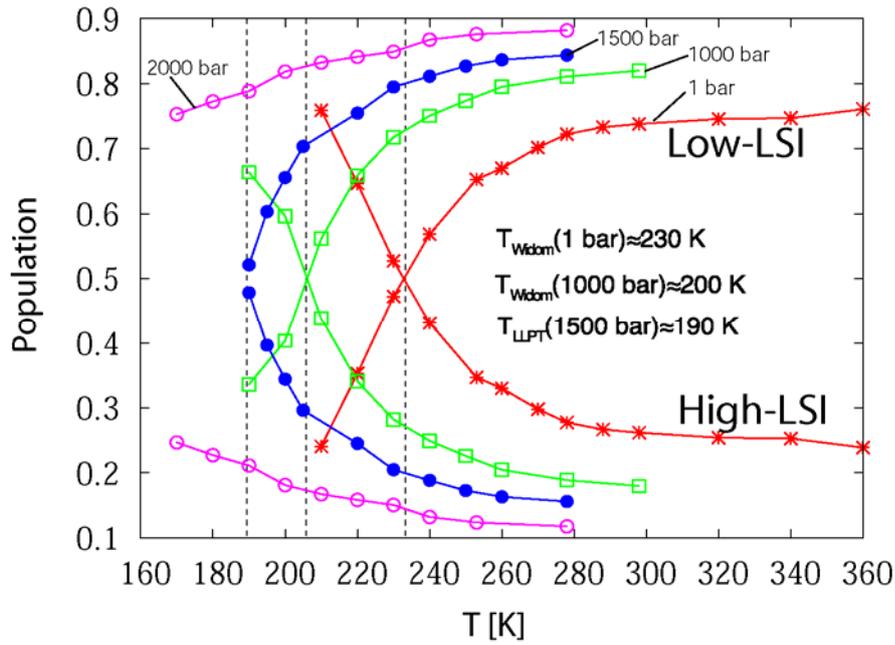

**Fig. 2.** Temperature dependence of the relative populations of high-*LSI* and low-*LSI* species in the inherent structures, defined according to the isosbestic minimum. The low-*LSI* and high-*LSI* curves are by definition mirror symmetric around the horizontal line. The temperature where a 1:1 ratio between low-*LSI* and high-*LSI* is realized coincides nearly perfectly with the Widom line at 1 and 1000 bar and the LLPT at 1500 bar. Note that at isothermal conditions, higher pressure greatly enhances the population of low-*LSI* species.

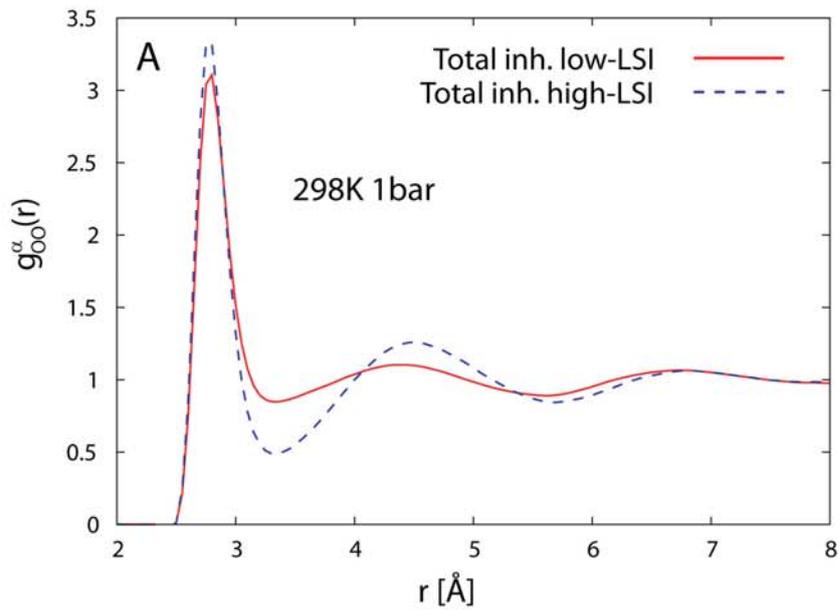

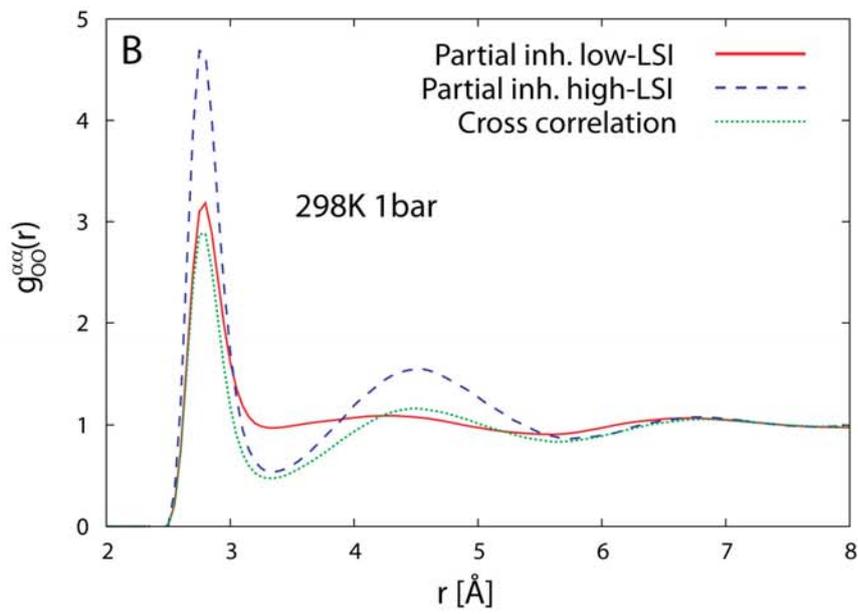

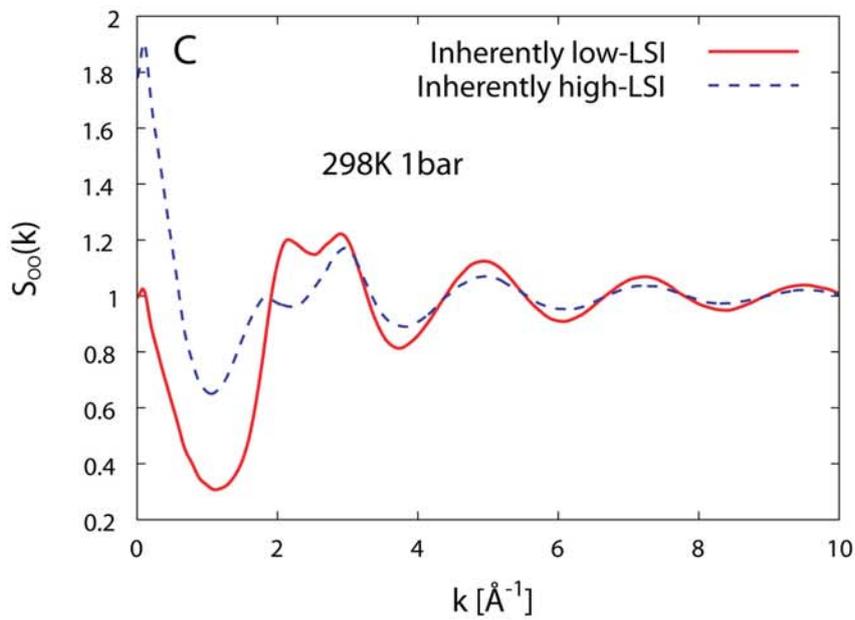

**Fig. 3. A:** Oxygen-oxygen pair-correlation function for subspecies that in the inherent structure (IS) belong to the low-*LSI* and high-*LSI* categories, respectively. Species with high *LSI* in the IS feature a sharp first peak and deep first minimum. Species with low *LSI* in the IS have a very shallow first minimum and poorly defined second shell, and the third shell is shifted to lower distances similarly to an experimental estimate of the PCF of high-density liquid water [16]. **B:** *Partial* pair-correlation functions involving either only high-*LSI* (blue, dashed) or low-*LSI* (red, full line) species and their cross correlation (green, full line). **C:** Partial structure factors for the same high-*LSI* and low-*LSI* subspecies. Note that the number density of the respective species have been used in calculating $S(k)$, explaining the stronger amplitude for the low-*LSI* species which constitute around 70% of all molecules at 298 K and 1 bar (see fig.2). Pronounced small-angle scattering is observed for both subspecies, directly evidencing their spatial separation.

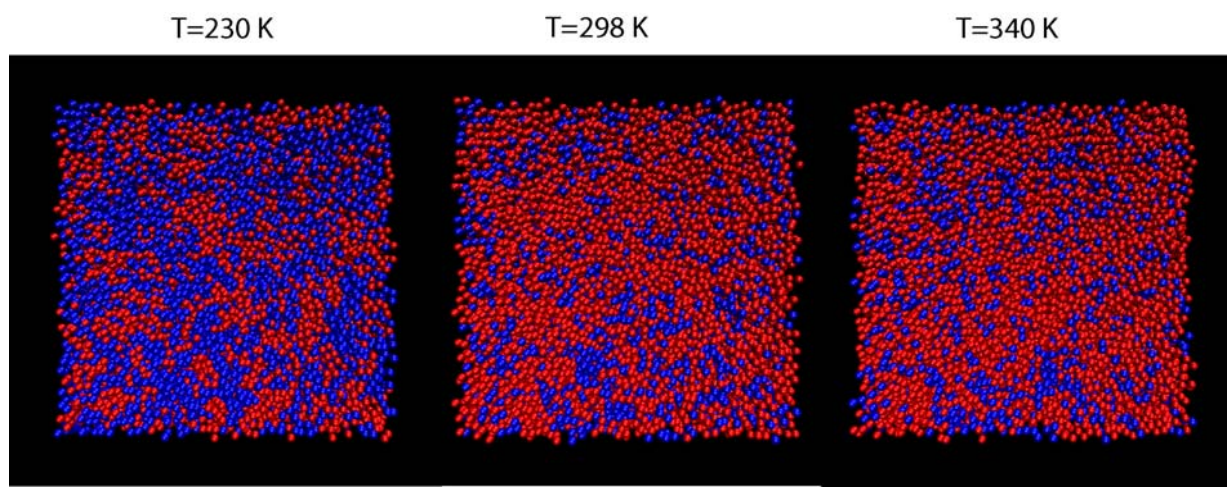

**Fig. 4:** Snapshots from the simulations at different temperatures (real structures, not energy minimized) with inherently low-LSI molecules colored red while inherently high-LSI species are blue. The largest clustering effect is seen at the Widom line (T=230 K), but clustering effects are seen also at 298 and 340 K. Visual comparison to snapshots with completely randomly classified subspecies confirms that the "clustering illusion" is not at work.

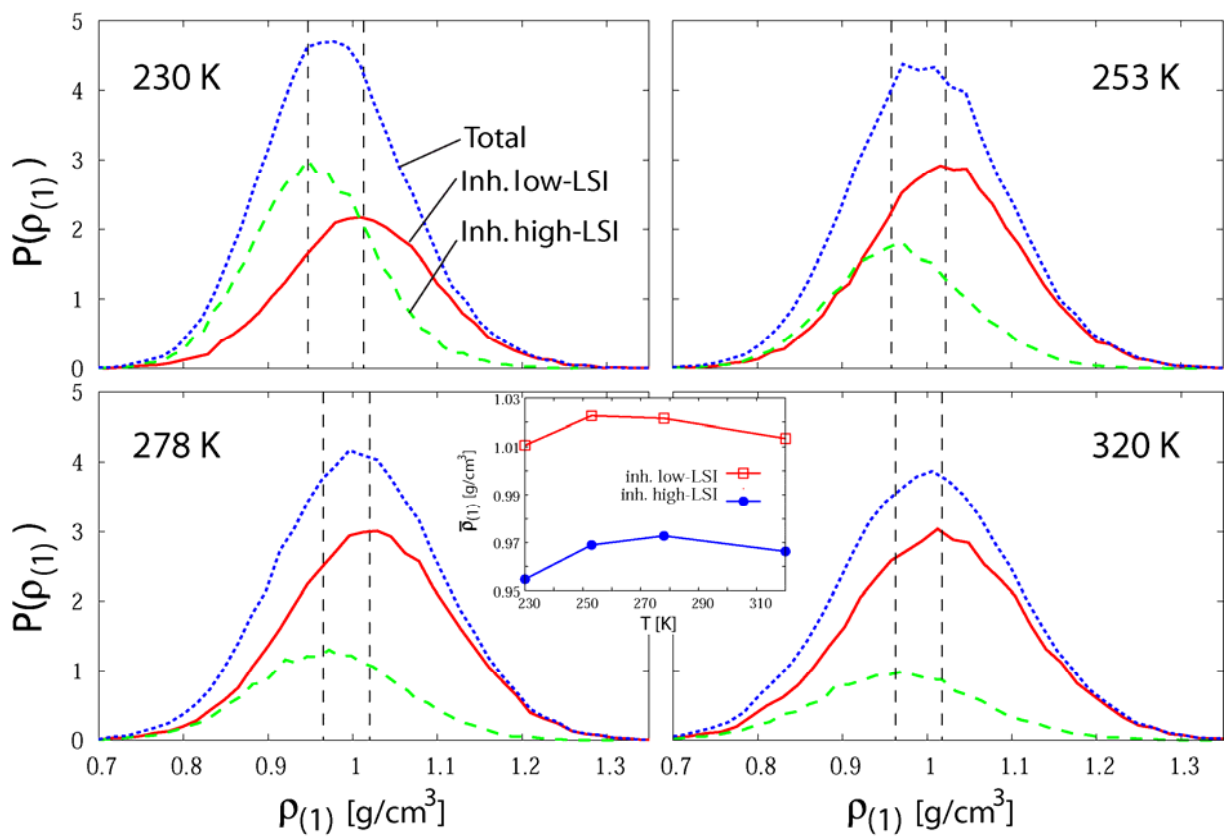

**Fig. 5.** Distributions of one-molecule local densities calculated using Voronoi polyhedra. Partial contributions from inherently low- and high-LSI species have been separated. The dashed vertical lines indicate the first-moments of the partial distributions, giving a mean density difference between the two partial distributions decreasing from 0.056 g/cm$^3$ at 230 K to 0.047 g/cm$^3$ at 320 K. The inset shows the mean densities of both components as function of temperature.